\pdfoutput=1

\documentclass[11pt,twoside,a4paper,cmspaper,final,collab]{cms-tdr}
\def\svnVersion{319214}\def\cmsCernNo{2013-037}\def\cmsCernDate{\today}\def\cmsMessage{Published in Physics Letters B as \href{http://dx.doi.org/10.1016/j.physletb.2016.01.010}{\doi{10.1016/j.physletb.2016.01.010.}}}

\begin{document}\cmsNoteHeader{HIN-14-007}

\hyphenation{had-ron-i-za-tion}
\hyphenation{cal-or-i-me-ter}
\hyphenation{de-vices}
\RCS$Revision: 319042 $
\RCS$HeadURL: svn+ssh://svn.cern.ch/reps/tdr2/papers/HIN-14-007/trunk/HIN-14-007.tex $
\RCS$Id: HIN-14-007.tex 319042 2016-01-21 21:04:56Z kjung $
\newlength\cmsFigWidth
\ifthenelse{\boolean{cms@external}}{\setlength\cmsFigWidth{0.85\columnwidth}}{\setlength\cmsFigWidth{0.4\textwidth}}
\ifthenelse{\boolean{cms@external}}{\providecommand{\cmsLeft}{top}}{\providecommand{\cmsLeft}{left}}
\ifthenelse{\boolean{cms@external}}{\providecommand{\cmsRight}{bottom}}{\providecommand{\cmsRight}{right}}
\providecommand{\rootsNN}{\ensuremath{\sqrt{s_{_\mathrm{NN}}}}}

\cmsNoteHeader{HIN-14-007}
\title{Transverse momentum spectra of inclusive b jets in pPb collisions at $\rootsNN = 5.02$\TeV}

\author*[purdue]{Kurt Jung}

\date{\today}

\abstract{
We present a measurement of b jet transverse momentum (\pt) spectra in proton-lead (pPb) collisions using a dataset corresponding to about 35\nbinv collected with the CMS detector at the LHC.  Jets from b quark fragmentation are found by exploiting the long lifetime of hadrons containing a b quark through tagging methods using distributions of the secondary vertex mass and displacement. Extracted cross sections for b jets are scaled by the effective number of nucleon-nucleon collisions and are compared to a reference obtained from \PYTHIA simulations of pp collisions. The \PYTHIA-based estimate of the nuclear modification factor is found to be $1.22\pm0.15\,(\text{stat+syst pPb})\pm0.27\,\text{(syst \PYTHIA)}$ averaged over all jets with \pt between 55 and 400\GeVc and with $\abs{\eta_{\text{lab}}}<2$.  We also compare this result to predictions from models using perturbative calculations in quantum chromodynamics.
}

\hypersetup{%
pdfauthor={CMS Collaboration},%
pdftitle={Transverse momentum spectra of inclusive b jets in pPb collisions at sqrt(s[NN]) = 5.02 TeV},%
pdfsubject={CMS},%
pdfkeywords={CMS, physics, heavy ions, b-tagging}
}

\newcommand{\etaCM}{\ensuremath{\eta_{\mathrm{CM}}}\xspace}
\newcommand{\etalab}{\ensuremath{\eta_{\text{lab}}}\xspace}
\newcommand{\RpAPythia}{\ensuremath{R_{\mathrm{pA}}^{\text{PYTHIA}}}\xspace}

\maketitle

\section{Introduction}
By colliding heavy nuclei at ultra-relativistic energies, sufficiently large energy densities are reached to form a quark-gluon plasma (QGP), a state which is characterized by an effective deconfinement of quarks and gluons~\cite{HIN-11-003,Adler2013}. Hard-scattered partons have been predicted to suffer energy loss as they traverse the QGP, primarily via collisional and radiative processes~\cite{Bjorken:1982tu,djordjevic2}.
This energy loss is commonly thought to be the mechanism responsible for the observed suppression of high transverse momentum (\pt) hadrons and jets in nucleus-nucleus collisions relative to proton-proton (pp) collisions~\cite{CasalderreySolana:2007zz,d'Enterria:2009am}. This suppression phenomenon, otherwise known as ``jet quenching'', was discovered at the RHIC experiments at BNL~\cite{Adcox:2004mh,Adams:2005dq,Back:2004je,Arsene:2004fa, Adams:2003im,Adams:2005ph, Adler:2003qi, Adare:2008ae} and has been investigated further using fully reconstructed jets at the CERN LHC~\cite{HIN-10-004,HIN-12-003,ATLASJetQ,ALICEJetQ}.
Studies of parton energy loss are expected to reveal the fundamental properties of the QGP.

The quenching of jets in heavy ion collisions should depend on the flavor of the fragmenting parton~\cite{CasalderreySolana:2007zz}. For example, under the assumption that radiative energy loss is the dominant mechanism, gluon jets are expected to quench more strongly than quark jets, owing to the larger color factor for gluon emission from gluons than from quarks~\cite{djordjevicGluon}.  There are also theoretical predictions that radiative energy loss may not be dominant for heavy quarks, including models based on collisional energy loss of quarks within the medium and models favoring an interpretation based on mesonic recombination and disassociation within the medium, \eg Refs.~\cite{vanHees:2005aa,teaney:2005aa}.  It is expected that there should be some mass-dependence of partonic energy loss at low momentum, and therefore, b quark jet (b jet) energy loss might be different from that of light quark jets~\cite{buzzatti:2011,djordevic:2015}. At high-\pt, however, the CMS Collaboration has shown that b jet suppression in PbPb is consistent with that of light quark jets above 80\GeVc~\cite{HIN-12-003}.

Here we present the first measurement of inclusive b-tagged jets in proton-lead (pPb) collisions.  This measurement in pPb provides the first direct evidence that the jet quenching observed in PbPb is dominated by final-state effects, rather than potential nuclear initial-state effects. Furthermore, these measurements will provide a factorization of cold nuclear matter effects from the medium suppression effects for jets in PbPb collisions.
Such a differentiation between initial-state and quenching effects as a function of flavor can place constraints on the energy loss mechanisms of partons in the hot and dense medium. This is especially important in light of the CMS measurement of the nuclear modification factor of charged particles in pPb collisions, which indicates surprisingly large initial-state effects \cite{HIN-12-017}. 

Measurements of dijets in pPb have also shown that a theoretical description of dijet yields as a function of pseudorapidity requires next-to-leading order effects with contributions from nuclear parton distribution functions (nPDFs)~\cite{HIN-13-001, ATLAS-CONF-2013-105}. While \PYTHIA simulations predict weak correlations of Bj{\"o}rken-x and single jet pseudorapidity, we investigate the pseudorapidity-dependent modification factor in order to probe for the presence of strong unanticipated effects in the heavy flavor sector. Any effects would depend predominantly on the gluon nPDFs, an area which has also been explored theoretically~\cite{ct10npdf}.  This is in contrast to the previous dijet measurement, where leading order quark jet processes have a significant contribution to the measurement, especially at high-\pt.

We present measurements of b jet production in pPb collisions at a nucleon-nucleon center-of-mass energy of $\rootsNN = 5.02$\TeV, recorded with the CMS detector, using an integrated luminosity of about 35\nbinv delivered by the LHC.  The cross section for b jets is measured and compared to pp cross sections simulated using the \PYTHIA event generator \cite{bib_pythia}, tune Z2~\cite{pythiaz2}. The resulting estimated nuclear modification factors (\RpAPythia) are compared to a prediction based on perturbative QCD (pQCD)~\cite{vitev:2013aa}.

\section{Detector and event selection}
The CMS detector has excellent capabilities to perform b jet identification (b tagging) as demonstrated in Ref.~\cite{Chatrchyan:2012jua}. The central feature of the CMS apparatus is a superconducting solenoid of 6\unit{m} internal diameter, providing a magnetic field of 3.8\unit{T}. Within the solenoid volume are a silicon pixel and strip tracker, a lead tungstate crystal electromagnetic calorimeter (ECAL), and a brass and scintillator hadron calorimeter (HCAL), each composed of a barrel and two endcap sections.
The tracker has a pseudorapidity coverage of $\abs{\etalab} < 2.4$, while the calorimetry covers $\abs{\etalab} < 3$. Muons are measured in gas-ionization detectors embedded in the steel flux-return yoke outside the solenoid. Extensive forward calorimetry complements the coverage provided by the barrel and endcap detectors. A more detailed description of the CMS detector, together with a definition of the coordinate system used and the relevant kinematic variables, can be found in Ref.~\cite{Chatrchyan:2008zzk}.

The event selection is identical to previous pPb jet analyses~\cite{HIN-12-017, HIN-13-001}, and includes the reconstruction of a primary interaction vertex, a careful removal of any noise artifacts from the hadronic calorimeter, along with a requirement that the primary interaction vertex is within 15\unit{cm} of the nominal interaction point along the beam axis.

In this analysis a new trigger combination algorithm is used, allowing for the maximization of statistical precision over a very large range of jet \pt. Triggers with thresholds ranging from 20 through 100\GeVc are combined.  Except for the 100\GeVc trigger, these triggers are prescaled, meaning only a fraction of the total number of events are recorded. In each event, the jet with the maximum online raw jet \pt (\ie the largest jet \pt seen by any of the five triggers) is found. If the highest-threshold trigger that should have identified the jet is absent (prescaled away), the whole event is rejected. Otherwise, all jets in the event are assigned a weight based on the prescale value of that highest-threshold trigger.  The resulting spectrum is fully efficient above $\approx$30\GeVc.

As described, \eg in Ref.~\cite{HIN-13-001}, the difference in the charge-to-mass ratio of protons and lead nuclei results in asymmetric beam energies for the two colliding species, which leads to a rapidity shift of 0.465 units between the nucleon-nucleon center-of-mass frame and the laboratory frame. In addition, after the data corresponding to an integrated luminosity of 20.9\nbinv were collected, the circulation directions of the proton and lead beams were reversed. This analysis will use \etaCM for the center-of-mass frame and \etalab for the lab frame pseudorapidities, where positive $\eta$ will always refer to the beam orientation where the proton beam direction is toward positive $z$.  In this orientation, $\etalab = \etaCM + 0.465$.

This analysis requires that all jets must have $-2.5 < \etaCM< 1.5$, which ensures that all jets fragment primarily within the tracker acceptance of $\abs{\etalab}<2.4$.
Finally, the background energy from the underlying pPb event is estimated in narrow ranges of pseudorapidity, as described in Ref.~\cite{Kodolova:2007hd}, and is subtracted from the jet. After the underlying event subtraction, jets must have a reconstructed $\pt> 55$\GeVc and a raw transverse momentum (before jet energy corrections) greater than 25\GeVc and must be found in an event where a single-jet trigger fires. This requirement is made in order to properly merge events from multiple triggers, as discussed earlier in this section.

In order to estimate the kinematic and resolution properties of jets, simulated dijet events are generated with \PYTHIA version 6.424, tune Z2 \cite{bib_pythia}.
These dijets are then embedded into a minimum bias pPb background event simulated by the \HIJING heavy ion event generator, version 1.383 \cite{bib_hijing}.

\section{Analysis procedure}

\subsection{Jet reconstruction}
Jets are reconstructed offline primarily from the energy deposits in the calorimeter towers, clustered by the anti-\kt algorithm~\cite{Cacciari:2008gp, Cacciari:2011ma} with a size parameter of 0.3. The constituent particles of the jet are reconstructed using the particle flow event algorithm, which identifies each individual particle with an optimized combination of information from the various elements of the CMS detector~\cite{CMS-PAS-PFT-09-001}. The raw jet energy is obtained from the sum of the tower energies, and the raw jet momentum by the vectorial sum of the constituent particle momenta, which results in a nonzero jet mass. The raw jet energies are then corrected to establish a uniform response of the calorimeter in $\eta$ and a calibrated absolute response in \pt. The final particle-flow-based jet energy resolution amounts typically to 15\% at 10\GeV, 8\% at 100\GeV, and 4\% at 1\TeV, to be compared to about 40\%, 12\%, and 5\% obtained when the calorimeters alone are used for jet clustering.

Jet energy corrections are derived from simulation, and are confirmed with in situ measurements of the energy balance in dijet and photon+jet events. Jet momentum is found from simulation to be within 1\% to 2\% of the true jet momentum over the whole \pt spectrum and detector acceptance used in this analysis. Additional selection criteria are applied to each event to remove spurious jet-like features originating from isolated noise patterns in certain HCAL regions.

\subsection{Tagging b jets}
Identification of b jets is based on kinematic variables related to the relatively long lifetime and large mass of B hadrons. Charged tracks associated with jets are used to reconstruct secondary vertices from B hadron and/or subsequent charm hadron decays from the b $\rightarrow$ c cascade. The primary discriminator used in this analysis to identify b jets takes advantage of the displaced secondary vertex. This secondary vertex based algorithm is called the ``simple secondary vertex" (SSV) tagger and is described in detail in Ref.~\cite{Chatrchyan:2012jua}.  Effectively, jets are assigned a discriminator value based on the secondary vertex flight distance significance, which is the ratio of the distance between the primary and secondary vertex to its uncertainty.  To remove additional contributions of jets from long lived light mesons, secondary vertex masses compatible with the $\PKzS$ meson and displacements larger than 2.5\unit{cm} are explicitly rejected. Using this discriminator, the contribution of b jets is enhanced by requiring that secondary vertices are far from the primary vertex. The SSV selection value used in this analysis is 2.0, requiring that the secondary vertex is two standard deviations away from the primary vertex.  This is chosen to give a misidentification rate on the order of 1\% for light-flavor jets and 10\% for charm jets, based on simulation. The corresponding b tagging efficiency is about 65\% for both pp and pPb collisions, which use identical reconstruction procedures.  This is in contrast to the PbPb b jet analysis at CMS where the b tagging efficiency is about 45\% due to the need for a dedicated regional track reconstruction owing to the very large multiplicities reached in central collisions \cite{HIN-12-003}.

\begin{figure}[htbp]
\centering
\includegraphics[width=0.49\textwidth]{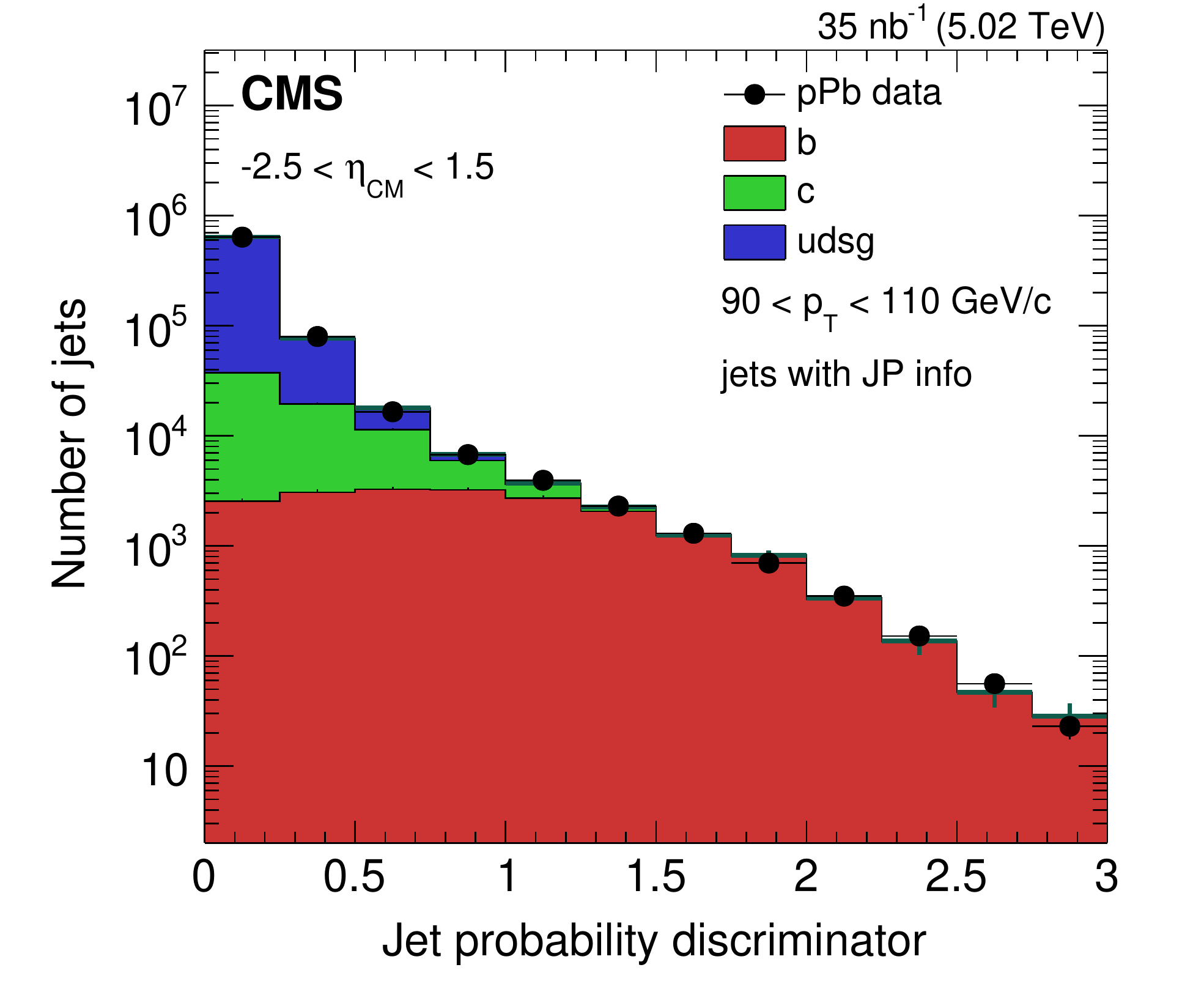}
\includegraphics[width=0.49\textwidth]{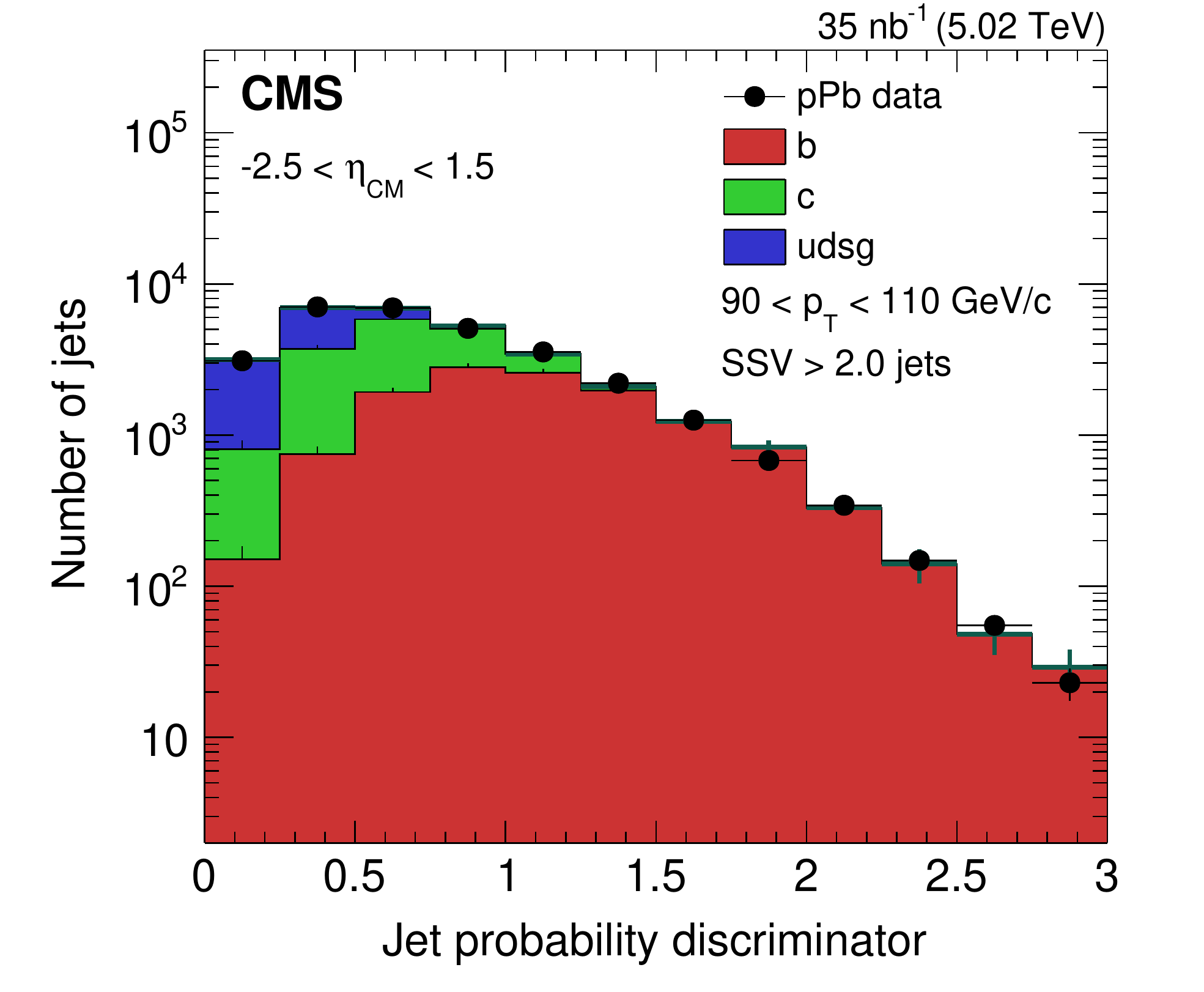}
\caption{\label{fig:JPdistr}
Distributions of the JP tagger discriminator before (\cmsLeft) and after (\cmsRight) applying the SSV tagger selection. Filled black points are data, while the colored histograms denote contributions from simulated b, c, and light-flavor jets in red, green and blue, respectively, obtained from a fit to data. Statistical uncertainties from data are in black, while statistical uncertainty from the templates are shown in dark green.
}
\end{figure}

The b tagging efficiency is obtained by simply counting the numbers of b jets before and after tagging in simulation, but is cross-checked using a data-driven method from the output of a second b tagging algorithm: the jet probability (JP) algorithm.
The advantage of this second tagger is that it does not rely upon the reconstruction of a secondary vertex \cite{Chatrchyan:2012jua}.  Instead, the JP tagger calculates the compatibility of each track in the jet cone with the primary vertex using a probability distribution based on the three-dimensional impact parameter significance.  In essence, the less compatible the jet tracks are with the primary vertex, the greater the likelihood of the jet being from a b quark fragmentation.  Tracks may also have a negative impact parameter, which arises when they are found to be displaced from the primary vertex on the opposite side of the vertex from the jet.  These tracks mainly come from primary particles with an improperly measured impact parameter due to finite vertex resolution effects or from poorly measured track kinematic parameters.  Since these types of tracks are essentially randomly associated with the vertex, they are not used to tag jets, but instead can be used to calibrate the tagger.  Randomly associated tracks should have no correlation to the vertex as a function of displacement, so the total distribution of these tracks as a function of track displacement should be flat.  If it is not, the tagger is calibrated by applying a weighting function in order to flatten the spectrum \cite{Chatrchyan:2012jua}.
Once the JP tagger is calibrated, discriminator values are obtained by calculating the sum of the
negative logarithm of all track-to-vertex probabilities, normalized by the factorial of the number of tracks associated with the jet.

\begin{figure}[htbp]
\centering
\includegraphics[width=0.49\textwidth]{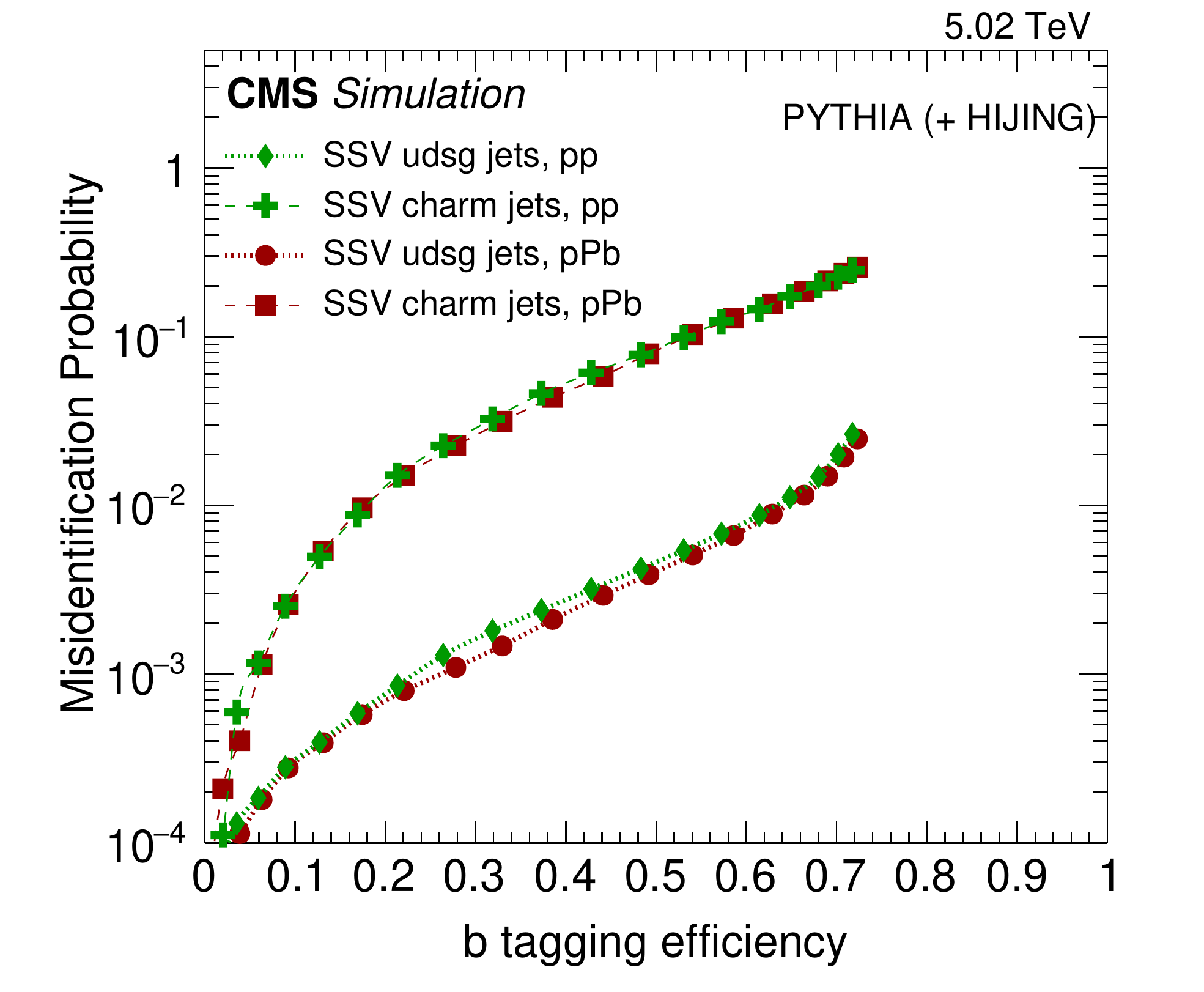}
\includegraphics[width=0.49\textwidth]{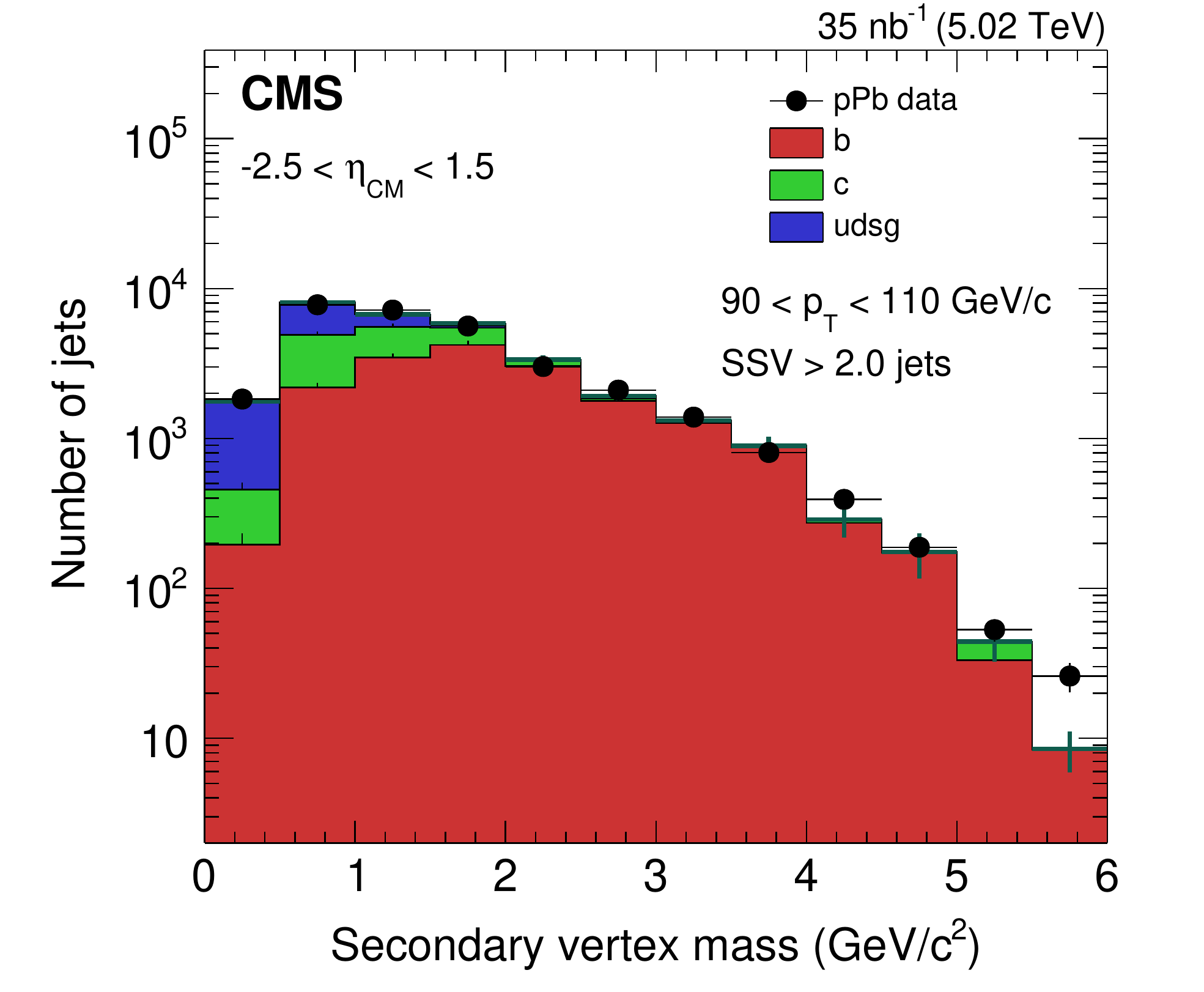} \caption{\label{fig:eff_vs_mistag}
The \cmsLeft~panel shows the likelihood of misidentifying a light-flavor (circles and dotted lines) or charm (squares and dashed lines) jet as a b jet, as a function of the b tagging efficiency.  Shown is the SSV tagger for pPb (purple) and pp (green) collisions.  The \cmsRight~panel shows a template fit to the secondary vertex invariant mass distribution in pPb collisions for jets with $90 < \pt < 110$\GeVc. Filled black points are data, while the colored histograms denote distributions of b, c, and light-quark jets in red, green and blue, respectively, extracted from the fit to data. Statistical uncertainties from data are shown as black vertical bars, while statistical uncertainties from the templates are shown as dark green vertical bars around the sum of the templates.
}
\end{figure}

Distributions of the JP tagger discriminator are plotted before and after applying the SSV selection defined earlier.  By using an unbinned maximum likelihood fit to the JP distributions, the three flavor contributions from simulations are simultaneously fit to the data.  From these fits, the SSV b tagging efficiency can be extracted based on Eq.~(\ref{eq:JPeff}), where $C_b$ is the fraction of b jets from simulation that have a JP discriminator value, $f_b^{\text{tagged}}$ is the purity of the SSV $>$2 tagged sample, $f_b^{\text{untagged}}$ is the purity before tagging, and $N_{\text{jets}}^{\text{untagged}}$ and $N_{\text{jets}}^{\text{tagged}}$ are the number of jets before and after the SSV selection, respectively. 
\begin{equation}
\epsilon_{SSV} = \frac{C_{b}f_{b}^{\text{tagged}}N_{\text{jets}}^{\text{tagged}}}{f_{b}^{\text{untagged}}{N_{\text{jets}}^{\text{untagged}}}}
\label{eq:JPeff}
\end{equation}

Example distributions of the JP tagger discriminator before and after SSV tagging in the range $90 < \pt <110$\GeVc are shown in Fig.~\ref{fig:JPdistr}.  The efficiency found by applying the SSV tagger to JP-tagged events in data and calculating the efficiency directly from simulation are compatible to within 5--20\%, where the difference is taken as a systematic uncertainty.

The b tagging efficiency of the SSV tagger is shown as a function of the misidentification probability of light-flavor and charm jets on the left in Fig. \ref{fig:eff_vs_mistag}. The efficiency and purity of the taggers are very similar in pp and pPb collisions due to the identical reconstruction methodology used for both collision types.  Though the JP tagger has a higher b tagging efficiency than the SSV tagger due to the fact that the JP tagger does not require the existence of a secondary vertex, the SSV tagger is the primary method of b jet identification in this analysis for two reasons. First, the SSV tagger is more robust against light-flavor and charm jet background due to the secondary vertex requirement.  Second, the JP tagger can be calibrated against data, which is essential to providing a data-driven estimate of the b tagging efficiency, therefore the JP tagger is better suited as a reference than the SSV tagger.

For each jet \pt bin, the b jet purity is extracted via a template fit.  For each secondary vertex, an invariant mass is calculated using the individual track energies and momenta.  Then, secondary vertex mass distributions from light, charm, and b jets in the \PYTHIA{+}\HIJING simulation are fit to those in data.  The shapes of the different flavor components of the distributions are fixed via the Monte Carlo simulations (MC), but the relative normalizations of each component are allowed to float independently.  While all jet flavors have significant contributions, the b jet contribution to the secondary vertex mass dominates above about 2\GeVcc, allowing for an accurate fit to data.  An example of such fitting is shown on the right in Fig. \ref{fig:eff_vs_mistag}.

For each tagger, a b jet yield can be calculated for a given \pt bin:  $N_{\PQb} = N f_{\PQb} / \epsilon$, where $N_{\PQb}$ is the number of b-tagged jets, $f_{\PQb}$ is the purity of the sample, derived from the secondary vertex mass fits, and $\epsilon$ is the tagger efficiency, determined from simulation.  After tagging, the jet resolution effects on the b jet \pt spectra are unfolded using a singular value decomposition (SVD) matrix inversion procedure \cite{svdunfold}, as implemented in the {\sc RooUnfold} package \cite{roounfold}. The pPb spectra are normalized by the total integrated luminosity (35\nbinv) and divided by the mass number of lead ($\mathrm{A} = 208$), which is the effective enhancement of jet production due to geometrical effects from the heavier nuclei, as predicted by the Glauber model~\cite{glauber}.
\begin{equation}
\label{eq:rpa}
R_{\mathrm{pA}}^{\mathrm{PYTHIA}} = \frac{1}{A}\frac{\rd^{2}\sigma^{\mathrm{pA}}_{\text{jet}}/\rd p_{T}\,\rd\eta}{\rd^2\sigma^{\mathrm{PYTHIA}}_{\text{jet}}/\rd p_{T}\,\rd\eta}.
\end{equation}

The formula used to calculate the nuclear modification factor (\RpAPythia) is defined in Eq. (\ref{eq:rpa}). The \etaCM-dependent \RpAPythia is obtained by dividing the jet cross section in pPb (scaled with the lead ion mass, A) by the jet cross section obtained from a pp reference. As there is no pp data available at $\rootsNN = 5.02$\TeV, this reference is obtained from a \PYTHIA calculation ``$\sigma^{\mathrm{PYTHIA}}_{\text{jet}}$".

\section{Systematic uncertainties}
\label{sec:syst}

The systematic uncertainties of the pPb yield fall into four general categories: b tagging, jet reconstruction, and scaling uncertainties due to unfolding and the luminosity uncertainty.
The b tagging uncertainties have five primary subcomponents. The first source of uncertainty comes from the difference between calculating the efficiency ($\epsilon$) using the JP tagger (Eq. (\ref{eq:JPeff}))~\cite{Chatrchyan:2012jua} and extracting $\epsilon$ directly from simulation. This is the dominant systematic uncertainty at high \pt and accounts for about 50\% of the total uncertainty.
A second source is obtained by varying the SSV tagger discriminator selection such that $\epsilon$ differs by about 10\%, which accounts for about 35\% of the total systematic uncertainty for jet \pt larger than about 100\GeVc and 10\% below 100\GeVc.
Next, the charm jet normalization is fixed to the light-flavor jet normalization, rather than allowing it to float independently in the template fits. This accounts for about 7\% of the total uncertainty and is independent of \pt.
Fourth, a data-derived (charm+light) background template produced from jets with small JP values is used.  This contribution is roughly 5\% of the total uncertainty for jet \pt larger than 100\GeVc and 50\% below 100\GeVc.
The final tagging uncertainty is found by varying the gluon splitting contribution in the b and c jet templates by 50\%.  This is the smallest contribution to the total systematic uncertainty (5\%). The total systematic uncertainty on the b jet tagging varies from about 15 to 20\% depending on the jet \pt. The uncertainty is evaluated via the quadratic sum of all systematic variations of the tagging procedure, which influence the extracted b tagging purity and efficiency values.

The jet reconstruction procedure has uncertainties totaling around 8--15\% for the pPb spectra stemming from closure tests between data and MC.  These uncertainties arise from the jet energy resolution (JER) and jet energy scale (JES).  The resolution uncertainty is about 10\%, which decreases as a function of jet \pt, while the scale uncertainty is about 3--4\%, depending on jet \pt.  The uncertainty stemming from the jet unfolding procedure is evaluated by varying the SVD regularization parameter and the presumed prior spectrum.  The pPb to pp normalization has about 5\% uncertainty due to the unfolding. Finally, the uncertainty on the pPb integrated luminosity is 3.6\% \cite{LUM-13-002}. These uncertainties are all summed in quadrature with the tagging uncertainties to obtain the total uncertainty on the pPb b jet spectra.

The pp reference cross section has two sources of systematic uncertainty.  As no pp data at 5\TeV exist yet, and since there are too few published measurements of b jet cross section to allow for an interpolated reference, we are forced to rely on simulation, but can make some reasonable assumptions regarding the expected agreement of the simulated reference with data.  These two sources of uncertainty are a 20\% uncertainty based on the discrepancy between existing b jet measurements and \PYTHIA simulations at 2.76~\cite{HIN-12-003} and 7\TeV~\cite{Chatrchyan:2012dk}, and a 8.5\% uncertainty based on the b jet cross section difference between the Z2 and D6T \cite{pythiaD6T} \PYTHIA tunes.
The discrepancies between \PYTHIA and data at 2.76 and 7\TeV are roughly constant in \pt and $\eta$, except for the \pt region well below the reach of this analysis, where the deviation becomes quite large. The data-to-simulation discrepancy is added in quadrature with the difference between the D6T and Z2 tune \pt distributions at both 2.76 and 7\TeV so that the difference in tune is accounted for in the overall pp uncertainty.
This 22\% overall pp uncertainty is shown as the red band around unity in Fig.~\ref{fig:spectra} (right panel) and in Fig.~\ref{fig:rpaVsPt}.

Lastly, the jet and b tagging systematic uncertainties for \RpAPythia are obtained by varying the pPb data simultaneously with the pp simulation in order to ensure any correlated systematics are cancelled out.  A partial cancellation of the uncertainties exists, but as the generator values are used for the pp reference in the analysis, the residual pPb unfolding uncertainties do not cancel, as would be the case with a pp measurement from data. It should also be noted that due to the template fitting procedure and unfolding, there is a partial correlation between the statistical and systematic uncertainties for the $\eta$-dependent result.

\section{Results}

\begin{figure}[htbp]
\centering
\includegraphics[width=0.49\textwidth]{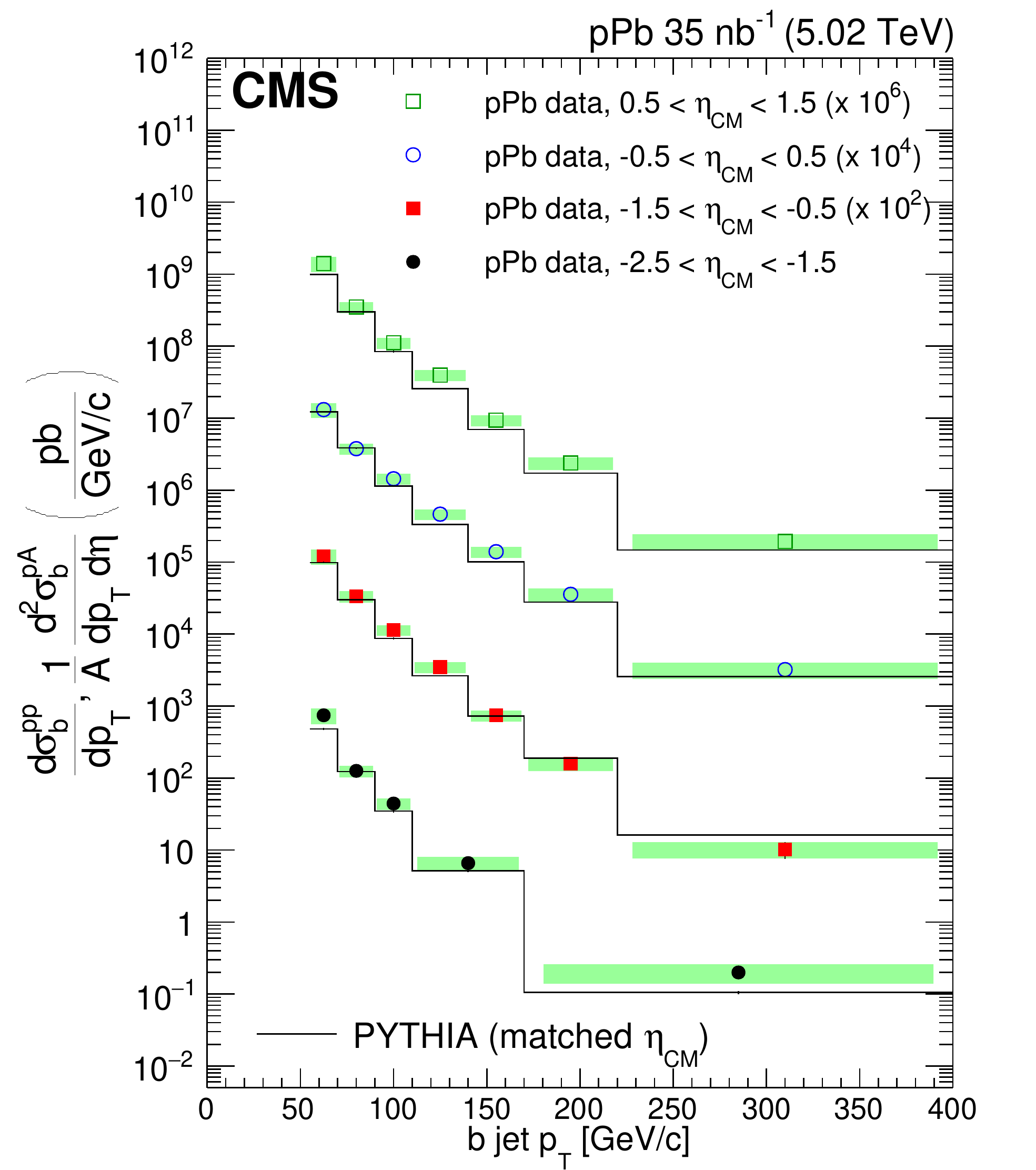}
\includegraphics[width=0.49\textwidth]{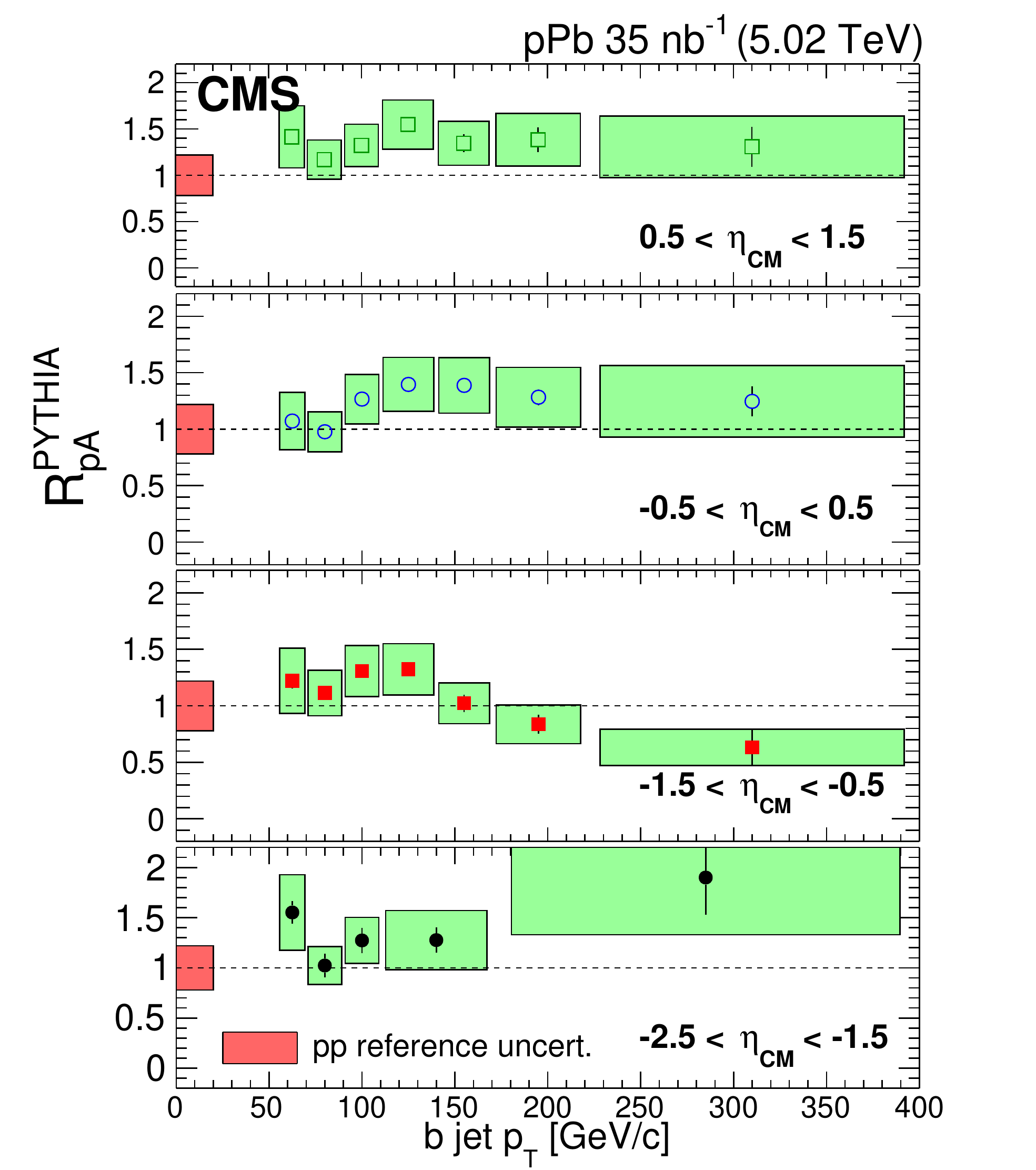} \caption[b jet spectra]{\label{fig:spectra}
The b jet cross section as a function of \pt is shown for various pseudorapidity selections for pPb collisions, scaled by the mass number of lead (filled boxes), and compared to \PYTHIA predictions of b jet cross sections in pp shown as bare histograms on the left. In addition, \RpAPythia\ measurements for the same four \etaCM ranges are shown on the right. Positive $\eta$ corresponds to the direction of the proton beam. Statistical uncertainties are represented using vertical bars, while systematic uncertainties are shown as colored bands on the left and filled boxes on the right. The pp reference uncertainties are shown separately as red boxes around unity on the right panel.
}
\end{figure}

The b jet \pt spectra in pPb are shown on the left in Fig.~\ref{fig:spectra} for several \etaCM selections, along with cross sections from the \PYTHIA pp reference (histograms).  We observe consistency between the pPb data and the \PYTHIA pp reference, indicating a lack of $\eta$-dependent effects.  This can be made explicit by calculating the \RpAPythia for each \etaCM selection, as defined in Eq.~(\ref{eq:rpa}).  The right side of Fig.~\ref{fig:spectra} shows the \RpAPythia measurements for the same four \etaCM selections as on the left. The average values are consistent with unity within uncertainties.

\begin{figure}[htbp]
\centering
  \includegraphics[width=0.49\textwidth]{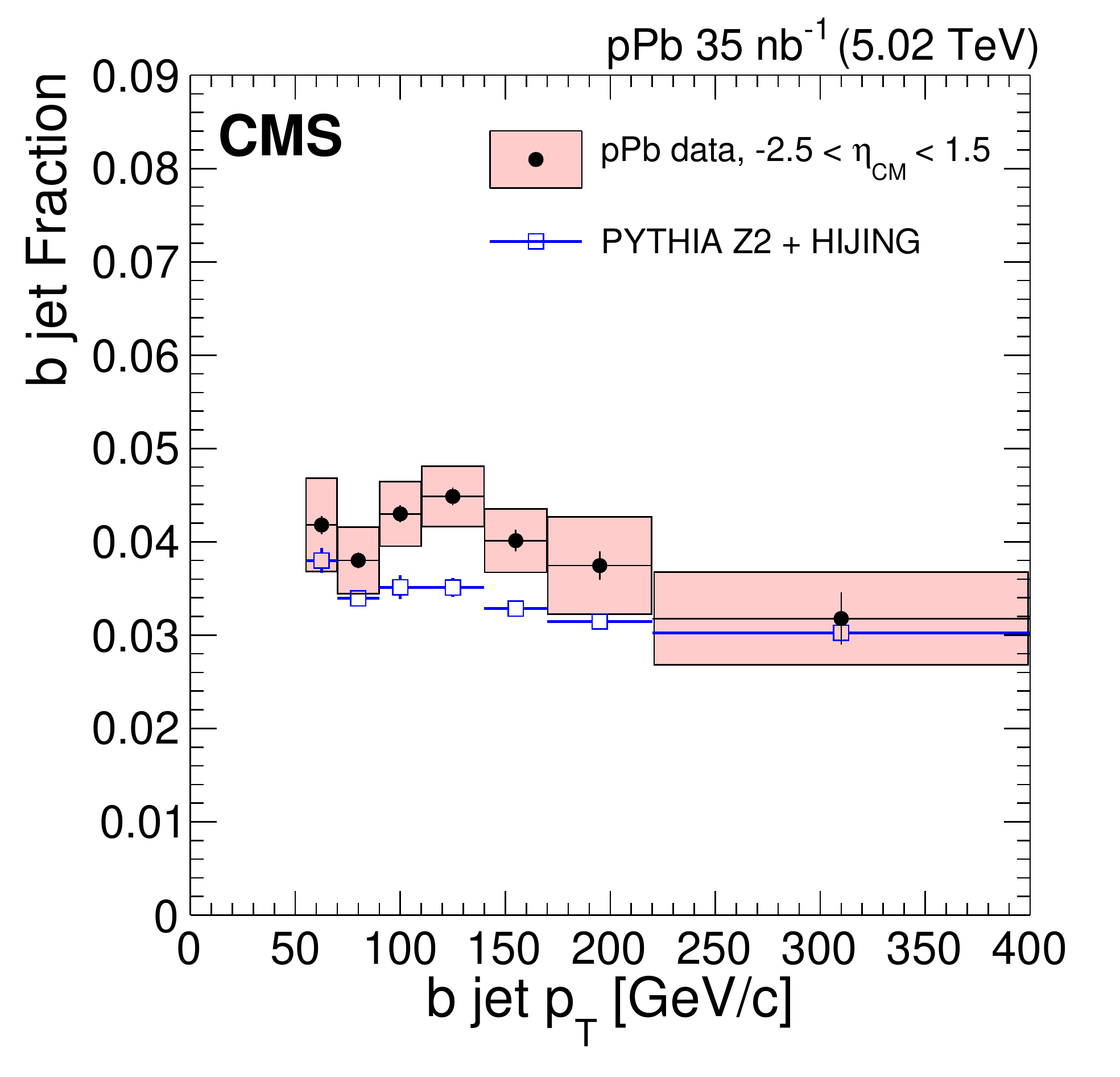}
  \caption[b jet fraction]{ The b jet fraction is shown for pPb data as filled black circles surrounded by filled boxes for systematic uncertainties, which are highly correlated to the b jet \RpAPythia uncertainties. A simulation from the Z2 tune of {\sc pythia+hijing} at $\rootsNN = 5.02$\TeV is also shown as open blue boxes.}
  \label{fig:bfraction}
  \includegraphics[width=0.49\textwidth]{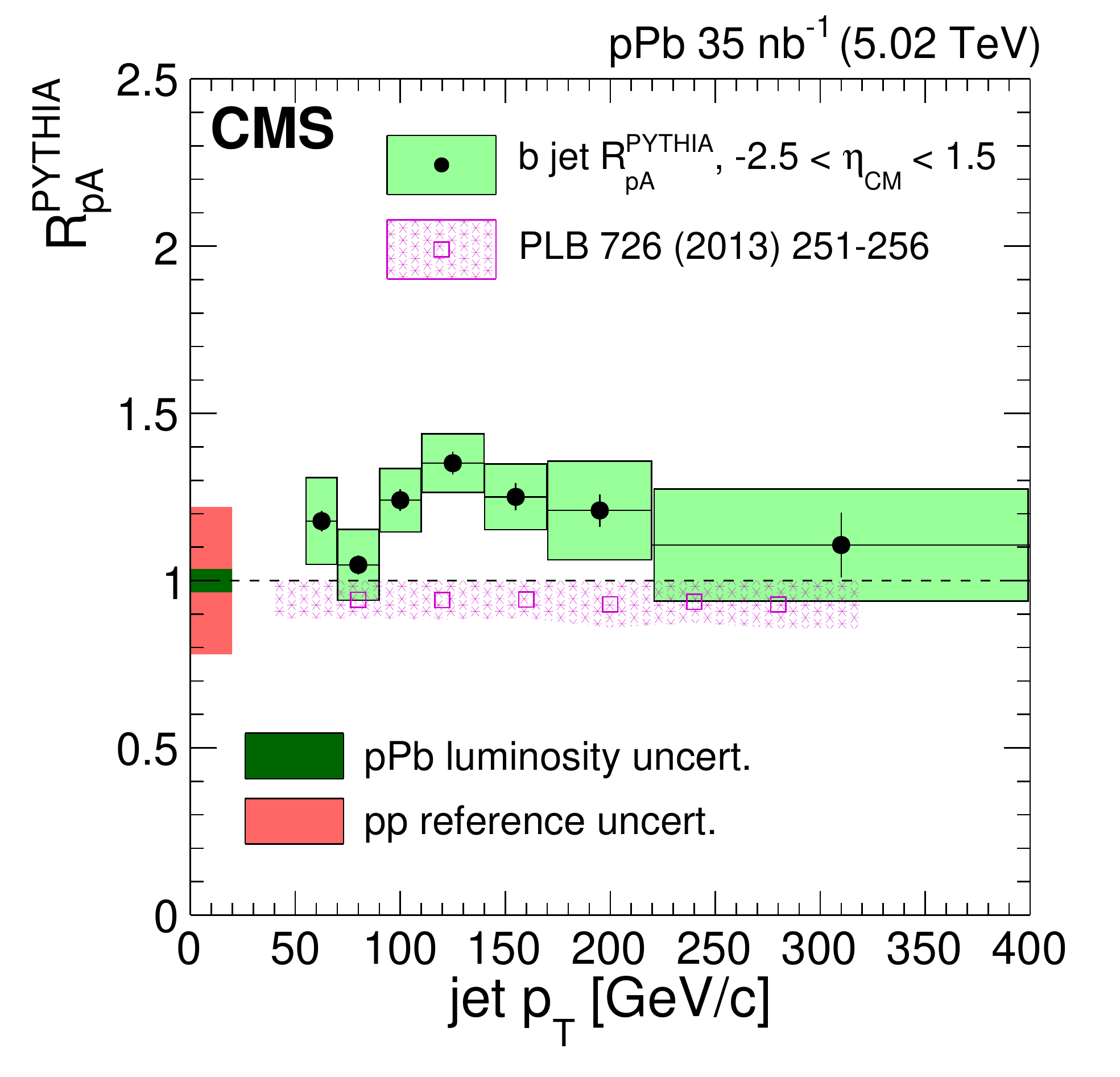}
  \caption[b jet $R_{\mathrm{pA}}$ vs. \pt]{The b jet \RpAPythia as a function of jet \pt is shown as points with filled boxes for systematic uncertainties.  The pp reference and integrated luminosity uncertainties are shown as red and green bands around unity, respectively. A pQCD prediction from Huang et. al. \cite{vitev:2013aa} is also shown.}
  \label{fig:rpaVsPt}
\end{figure}

The b jet fraction can be extracted by dividing the b jet cross section by the inclusive jet cross section.  This is shown in Fig.~\ref{fig:bfraction}, where we observe consistent results between the pPb data and the \PYTHIA simulation within systematic uncertainties.  These systematic uncertainties are calculated by noting that the uncertainties from the jet energy scale, unfolding procedure, and the luminosity are highly correlated between the samples with and without implementing b tagging, and we therefore assign the b tagging uncertainties as the total uncertainty on the fraction.

Figure~\ref{fig:rpaVsPt} shows the pseudorapidity-integrated \RpAPythia. Fitting a constant to this distribution returns a value of $\RpAPythia = 1.22 \pm 0.15\,(\text{stat}+\text{syst pPb})\pm0.27\,(\text{syst \PYTHIA})$, which indicates that the b jet yield in pPb is consistent with the pp \PYTHIA simulation, especially considering the 22\% uncertainty on just the \PYTHIA reference. The measurement does not, however, exclude an enhancement in $R_{\mathrm{pA}}$ as large as the one observed in the charged particle measurement from CMS at high \pt~\cite{HIN-12-017}.  In addition, Fig.~\ref{fig:rpaVsPt} shows the comparison of the measured \RpAPythia to predictions from a pQCD model that includes modest initial-state energy-loss effects \cite{vitev:2013aa} and conservative uncertainties stemming from the unknown fraction of jets that contain a collimated \bbbar pair originating from gluon splitting. The model and data are roughly consistent within the total systematic uncertainties from both \PYTHIA and the pPb data.

This result can be compared to the recent study of B meson production in pPb from the CMS Collaboration \cite{HIN-14-004}.  We find good agreement between the two analyses, noting that the b jet \RpAPythia value is consistent with the observed $R_{\mathrm{pA}}^{\mathrm{FONLL}}$ values for all B mesons over the entirety of the 10--60\GeVc \pt range used in the meson analysis.

\section{Conclusions}	
In summary, the first measurements of b jet production at 5.02\TeV have been presented over a \pt range from 55--400\GeVc and a pseudorapidity window of $-2.5 < \etaCM< 1.5$.  The observed value of $\RpAPythia = 1.22 \pm 0.15\,(\text{stat}+\text{syst pPb})\pm0.27\,(\text{syst \PYTHIA})$ provides the first direct evidence that cold nuclear matter effects do not play a major role in jet quenching in the PbPb system.  Furthermore, a sizeable jet production enhancement from cold nuclear matter effects is not expected at such large \pt, a conclusion which the data supports. We find that the pseudorapidity-integrated and pseudorapidity-dependent \RpAPythia values are consistent both with unity and with the enhancement observed by CMS for charged particles at high \pt.  

The consistency with unity as a function of pseudorapidity indicates that very large nPDF effects do not exist in the gluon sector, an observation which is consistent with theoretical models. While the current constraints on the gluon nPDFs are not very tight due to present systematic uncertainties, this measurement provides a baseline for future studies, especially those that use a true proton-proton sample as a reference point. Studies of back-to-back b jets, for example, will provide significant constraints on these nPDF effects due to the tighter correlation of pseudorapidity and Bj{\"o}rken-x and the restriction of b jet production to primarily leading order processes.

Overall, these results provide a baseline for the study of in-medium b quark energy loss in PbPb collisions. Future measurements of b jets in pp collisions at 5.02\TeV will reduce the large systematic uncertainties from the current \PYTHIA reference, allowing for a more precise measurement of b jet energy modification in pPb collisions.

\begin{acknowledgments}
We congratulate our colleagues in the CERN accelerator departments for the excellent performance of the LHC and thank the technical and administrative staffs at CERN and at other CMS institutes for their contributions to the success of the CMS effort. In addition, we gratefully acknowledge the computing centres and personnel of the Worldwide LHC Computing Grid for delivering so effectively the computing infrastructure essential to our analyses. Finally, we acknowledge the enduring support for the construction and operation of the LHC and the CMS detector provided by the following funding agencies: BMWFW and FWF (Austria); FNRS and FWO (Belgium); CNPq, CAPES, FAPERJ, and FAPESP (Brazil); MES (Bulgaria); CERN; CAS, MoST, and NSFC (China); COLCIENCIAS (Colombia); MSES and CSF (Croatia); RPF (Cyprus); MoER, ERC IUT and ERDF (Estonia); Academy of Finland, MEC, and HIP (Finland); CEA and CNRS/IN2P3 (France); BMBF, DFG, and HGF (Germany); GSRT (Greece); OTKA and NIH (Hungary); DAE and DST (India); IPM (Iran); SFI (Ireland); INFN (Italy); MSIP and NRF (Republic of Korea); LAS (Lithuania); MOE and UM (Malaysia); CINVESTAV, CONACYT, SEP, and UASLP-FAI (Mexico); MBIE (New Zealand); PAEC (Pakistan); MSHE and NSC (Poland); FCT (Portugal); JINR (Dubna); MON, RosAtom, RAS and RFBR (Russia); MESTD (Serbia); SEIDI and CPAN (Spain); Swiss Funding Agencies (Switzerland); MST (Taipei); ThEPCenter, IPST, STAR and NSTDA (Thailand); TUBITAK and TAEK (Turkey); NASU and SFFR (Ukraine); STFC (United Kingdom); DOE and NSF (USA).

Individuals have received support from the Marie-Curie programme and the European Research Council and EPLANET (European Union); the Leventis Foundation; the A. P. Sloan Foundation; the Alexander von Humboldt Foundation; the Belgian Federal Science Policy Office; the Fonds pour la Formation \`a la Recherche dans l'Industrie et dans l'Agriculture (FRIA-Belgium); the Agentschap voor Innovatie door Wetenschap en Technologie (IWT-Belgium); the Ministry of Education, Youth and Sports (MEYS) of the Czech Republic; the Council of Science and Industrial Research, India; the HOMING PLUS programme of the Foundation for Polish Science, cofinanced from European Union, Regional Development Fund; the OPUS programme of the National Science Center (Poland); the Compagnia di San Paolo (Torino); the Consorzio per la Fisica (Trieste); MIUR project 20108T4XTM (Italy); the Thalis and Aristeia programmes cofinanced by EU-ESF and the Greek NSRF; the National Priorities Research Program by Qatar National Research Fund; the Rachadapisek Sompot Fund for Postdoctoral Fellowship, Chulalongkorn University (Thailand); and the Welch Foundation, contract C-1845.
\end{acknowledgments}

\bibliography{auto_generated}

\cleardoublepage \appendix\section{The CMS Collaboration \label{app:collab}}\begin{sloppypar}\hyphenpenalty=5000\widowpenalty=500\clubpenalty=5000\textbf{Yerevan Physics Institute,  Yerevan,  Armenia}\\*[0pt]
V.~Khachatryan, A.M.~Sirunyan, A.~Tumasyan
\vskip\cmsinstskip
\textbf{Institut f\"{u}r Hochenergiephysik der OeAW,  Wien,  Austria}\\*[0pt]
W.~Adam, E.~Asilar, T.~Bergauer, J.~Brandstetter, E.~Brondolin, M.~Dragicevic, J.~Er\"{o}, M.~Flechl, M.~Friedl, R.~Fr\"{u}hwirth\cmsAuthorMark{1}, V.M.~Ghete, C.~Hartl, N.~H\"{o}rmann, J.~Hrubec, M.~Jeitler\cmsAuthorMark{1}, V.~Kn\"{u}nz, A.~K\"{o}nig, M.~Krammer\cmsAuthorMark{1}, I.~Kr\"{a}tschmer, D.~Liko, T.~Matsushita, I.~Mikulec, D.~Rabady\cmsAuthorMark{2}, B.~Rahbaran, H.~Rohringer, J.~Schieck\cmsAuthorMark{1}, R.~Sch\"{o}fbeck, J.~Strauss, W.~Treberer-Treberspurg, W.~Waltenberger, C.-E.~Wulz\cmsAuthorMark{1}
% ...
\vskip\cmsinstskip
% ...
1:~~Also at Vienna University of Technology, Vienna, Austria\\
2:~~Also at CERN, European Organization for Nuclear Research, Geneva, Switzerland\\
% ...

\end{sloppypar}
\end{document}